\documentclass[submission, Proceedings]{SciPost}

\hypersetup{
    colorlinks,
    linkcolor={red!50!black},
    citecolor={blue!50!black},
    urlcolor={blue!80!black}
}

\usepackage{subcaption}
\usepackage{lineno}
\usepackage[bitstream-charter]{mathdesign}
\DeclareSymbolFont{usualmathcal}{OMS}{cmsy}{m}{n}
\DeclareSymbolFontAlphabet{\mathcal}{usualmathcal}

\begin{document}

\begin{center}{\Large \textbf{
Recent LHCb results on forward physics and diffraction\\
}}\end{center}

\begin{center}
Luke McConnell\textsuperscript{1$\star$}
\end{center}

\begin{center}
{\bf 1} University College Dublin
\\
* lucas.mcconnell@cern.ch
\end{center}

\begin{center}
\today
\end{center}


\definecolor{palegray}{gray}{0.95}
\begin{center}
\colorbox{palegray}{
  \begin{tabular}{rr}
  \begin{minipage}{0.1\textwidth}
    \includegraphics[width=23mm]{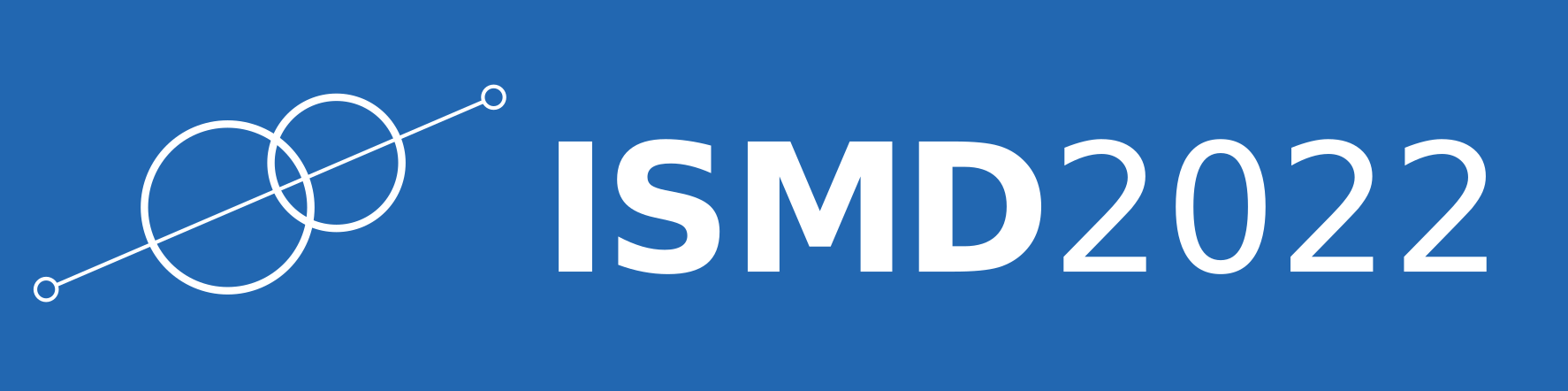}
  \end{minipage}
  &
  \begin{minipage}{0.8\textwidth}
    \begin{center}
    {\it 51st International Symposium on Multiparticle Dynamics (ISMD2022)}\\ 
    {\it Pitlochry, Scottish Highlands, 1-5 August 2022} \\
    \doi{10.21468/SciPostPhysProc.?}\\
    \end{center}
  \end{minipage}
\end{tabular}
}
\end{center}

\section*{Abstract}
{\bf
The LHCb detector at the LHC offers unique coverage of forward rapidities for studies of Central Exclusive Production (CEP) and soft QCD. CEP measurements allow the investigation of the nature of pomerons, and provide constraints on low-x gluon phenomenology, probing potential saturation effects. Moreover LHCb can test phenomenological models of soft QCD processes, by measuring the production of forward hadrons in pp collisions. In this talk the latest results from the LHCb experiment will be presented.
}


\section{Introduction}
\label{sec:intro}
In these proceedings, four recent LHCb papers are presented and the main results summarised. The first two papers involve Central Exclusive Production (CEP) \cite{Albrow:2010yb}, a class of diffractive \footnote{A diffractive process is a particle scattering with rapidity gaps present between the outgoing particles. The name is used in strict analogy with optical diffraction, as the differential cross section as a function of Mandelstam $t$ has a prominent forward peak and successive decreasing dips reminiscent of the intensity minima seen in optical diffraction.} processes with two rapidity gaps present on either side of the produced particles. The interaction is mediated by the exchange of a colourless object called the pomeron \cite{Kuraev:1977fs,Balitsky:1978ic}, which can be thought of as two or more gluons in a colour-singlet state. CEP occurs in ultra-peripheral collisions (UPCs), where the impact parameter is greater than the sum of the radii of the colliding hadrons. This results in the short range strong force being mitigated, and the colliding hadrons remain intact.

\section{Apparatus}
\label{sec:apparatus}
The LHCb detector \cite{LHCb:2008vvz} is a single arm forward spectrometer with an asymmetric design about the interaction point and a pseudorapidity range of $2 < \eta <5$. It has a high-precision tracking system, which includes a silicon-strip vertex detector, called the VELO, located upstream of the dipole magnet. Three stations of silicon-strip detectors and straw drift tubes are situated downstream of the magnet. Information from two ring-imaging Cherenkov detectors is used to differentiate charged hadrons. Photons, electrons, and hadrons are identified by the calorimetry system, which consists of scintillating-pads and pre-shower detectors, an electromagnetic, and a hadronic calorimeter. Muons are detected by the muon system which is made up of alternating layers of iron and multi-wire proportional chambers. The HeRSCheL \cite{Akiba:2018neu} detector is a set of scintillating tiles in the high rapidity regions. It is used to suppress contamination coming from inelastic events involving hadron dissociation, as opposed to in UPCs where the colliding hadrons remain intact.

\section{Central Exclusive Production of $J/\varPsi$ and $\psi(2S)$ mesons in $pp$ collisions at $\sqrt{s} = 13$ TeV}
\label{sec:paper1}
\begin{figure}[h]
\centering
\begin{subfigure}{.475\linewidth}
\includegraphics[width=1.0\linewidth]{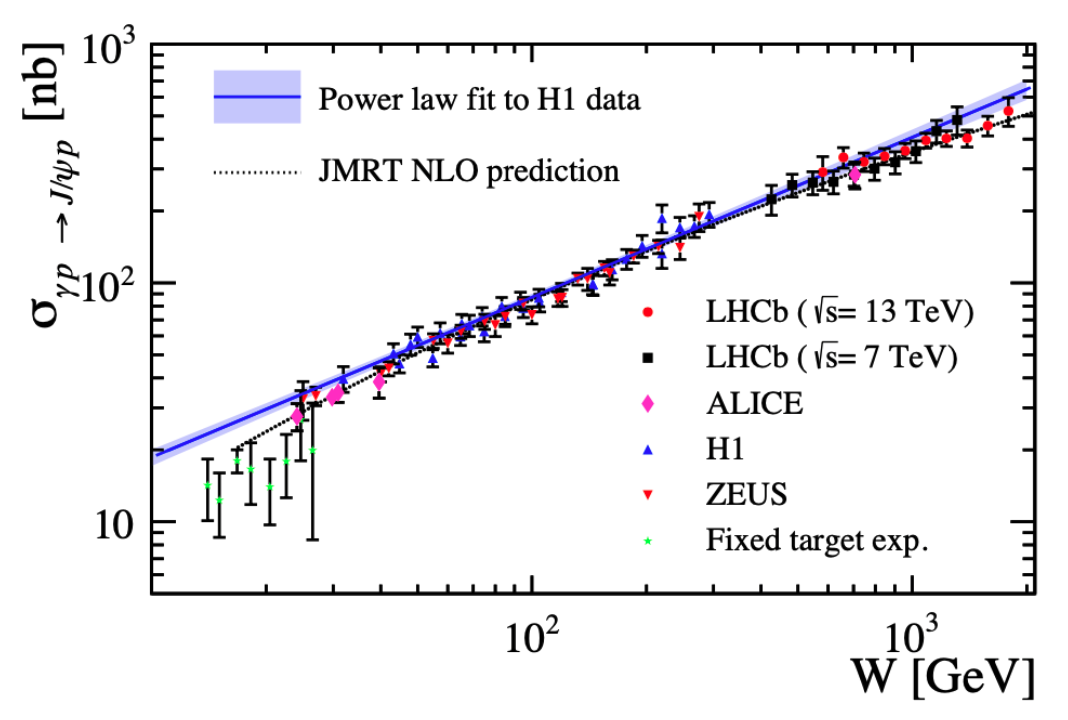}
\caption{$J/\Psi$}
\end{subfigure}
\begin{subfigure}{.485\linewidth}		   							   		\includegraphics[width=1.0\linewidth]{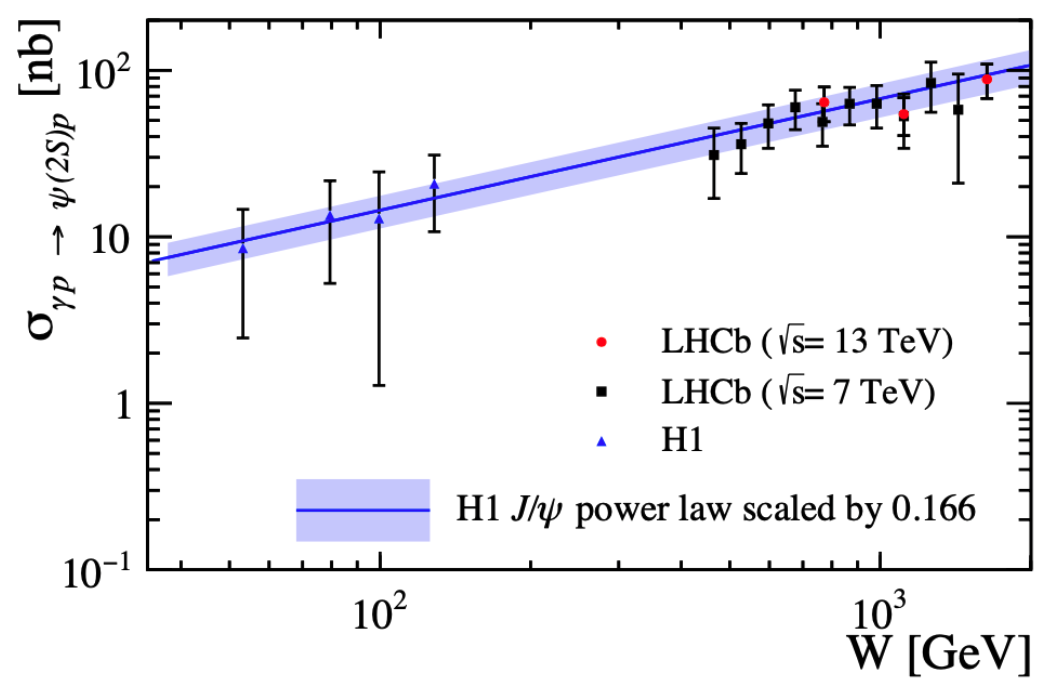}
\caption{$\psi(2S)$}
\end{subfigure}
\caption{Photo-production differential cross sections for the $J/\varPsi$ and the $\psi(2S)$ as a function of $W$. Also plotted are results from other measurements and NLO JMRT predictions. \cite{LHCb:2018rcm}}
\label{paper1}
\end{figure}

The measurements in this analysis \cite{LHCb:2018rcm} of CEP of the $J/\varPsi$ and $\psi(2S)$ mesons can help provide a test of perturbative-QCD \cite{Harland-Lang:2014lxa}, probe the pomeron \cite{Albrow:2010yb}, and constrain the gluon PDF \cite{Guzey:2013qza}. The signal process proceeds via the fusion of a photon and a pomeron, with the produced mesons subsequently decaying to the dimuon channel. This results in a clean experimental signature in contrast to an inelastic collision. Inelastic background is suppressed using information from the HeRSCheL detector. The plots in Figure \ref{paper1} show the measured photo-production differential cross sections for the $J/\varPsi$ and the $\psi(2S)$ mesons as a function of the photon-proton centre-of-mass energy $\left(W\right)$. These results are plotted alongside NLO JMRT predictions \cite{Jones:2013pga,Jones:2013eda} as well as prior results from H1 \cite{H1:2013okq}, ZEUS \cite{ZEUS:2002wfj}, ALICE \cite{ALICE:2012yye}, fixed target data \cite{PhysRevLett.48.73,PhysRevLett.52.795,1993197}, and the LHCb 7 TeV result \cite{LHCb:2014acg}. The 13 TeV result agrees with the 7 TeV result in the region of kinematic overlap. The cross section for the $J/\varPsi$ deviates at higher energies from a pure power-law extrapolation of the lower energy H1 data. The $\psi(2S)$ results are consistent but more data is required in the channel in order to make a critical comparison.
	
\section{Study of the coherent charmonium production in ultra-peripheral lead-lead collisions}
\label{sec:paper2}
\begin{figure}[h]
\centering
\includegraphics[width=1.\textwidth]{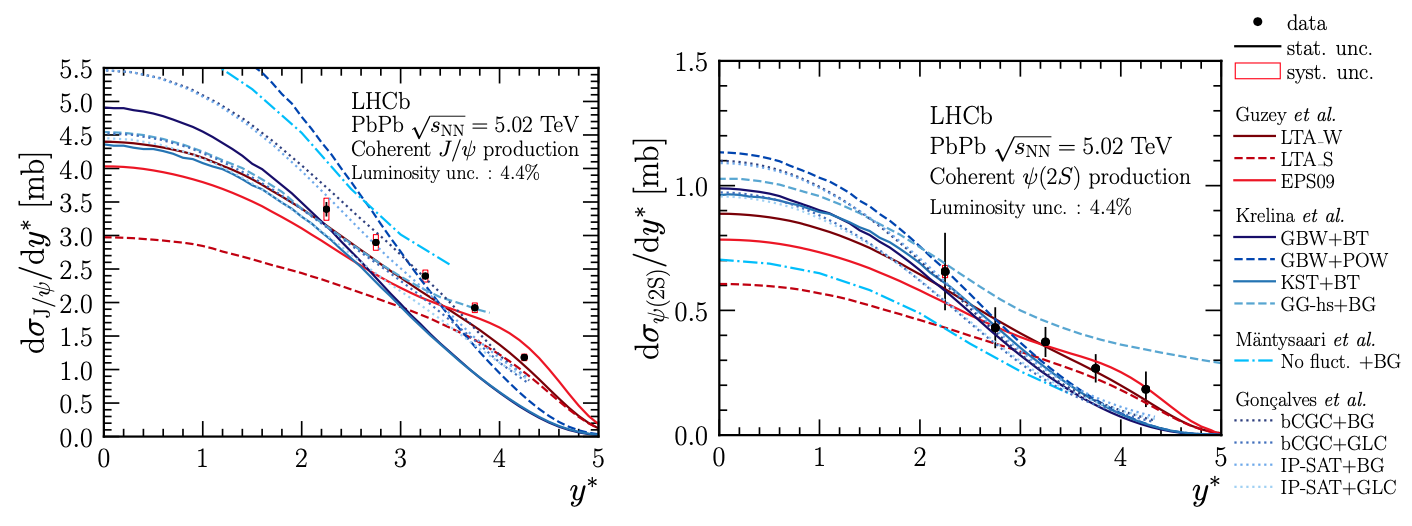}
\caption{Differential cross sections as function of rapidity for the $J/\varPsi$ and $\psi(2S)$ mesons. Also plotted are several model predictions. \cite{LHCb:2022ahs}}
\label{paper2}
\end{figure}

The photon flux from a hadron is proportional to the square of its electric charge, hence photon-nucleon interactions are significantly enhanced in $Pb-Pb$ collisions compared to $p-p$ collisions. In coherent production, the photon emitted from the ion interacts with a pomeron emitted by the entire nucleus. In incoherent production, the pomeron is emitted by a single nucleon within the nucleus. The measurement \cite{LHCb:2022ahs} allows for the study of nuclear shadowing effects \cite{AyalaFilho:2008zr} and of the initial states of collisions with small $x$ \cite{Jones:2015nna}. The signal decay is in the dimuon channel. First, the yields for the signal candidates and for the non-resonant background are determined from an unbinned fit to the dimuon invariant mass distributions. In the second step, fits to the $p_{T}$ distributions are used to determine the coherent yields. The plots in Figure \ref{paper2} show the differential cross sections as function of rapidity $\left(y^{*}\right)$ for the $J/\varPsi$ and $\psi(2S)$. Shown also are several theoretical predictions. Models which are based on perturbative-QCD (pQCD) using the leading-log approximation are in red \cite{Guzey:2016piu,Guzey:2016qwo}, while those in blue are based on colour-glass-condensate (CGC) calculations. The pQCD calculations are done with weak and strong leading twist nuclear shadowing models \cite{Frankfurt:2011cs}. These models are compatible with the data, with good agreement at high rapidity and a slight underestimation at low rapidity for both mesons. The blue CGC models include the ones by \emph{Krelina et al.} \cite{Cepila:2017nef,Kopeliovich:2020has}, which can be considered as variations of the colour-dipole model. The prediction by \emph{M\"{a}ntysaari et al.} \cite{Mantysaari:2017dwh,Lappi:2014eia} also uses the colour-dipole model and includes sub-nucleon scale fluctuations. The models by \emph{Gon\c{c}alves et al.} \cite{Goncalves:2017wgg,Goncalves:2005yr} depend on the dipole-hadron scattering amplitude and vector meson wave function. The models based on the CGC are compatible with the data, with large variations between them. Better agreement is seen at low rapidity than at high rapidity.

\section{$J/\varPsi$ photo-production in Pb-Pb peripheral collisions at $\sqrt{s_{NN}} = 5$ TeV}
\label{sec:paper3}
\begin{figure}[h]
\centering
\begin{subfigure}{.475\linewidth}
\includegraphics[width=1.0\linewidth]{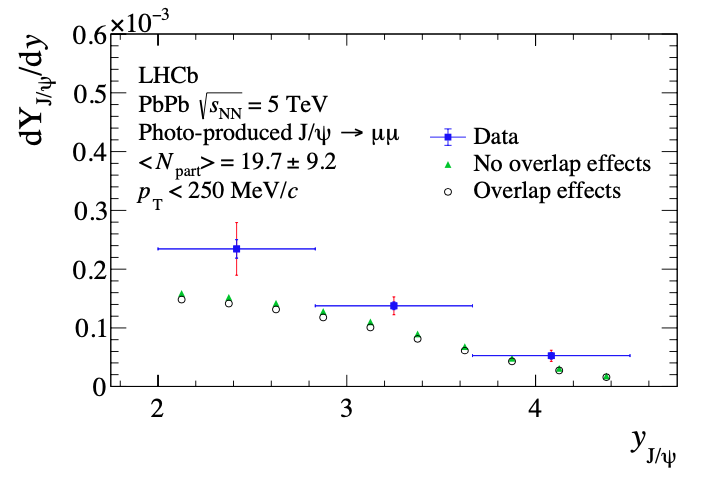}
\end{subfigure}
\begin{subfigure}{.485\linewidth}		   							   		\includegraphics[width=1.0\linewidth]{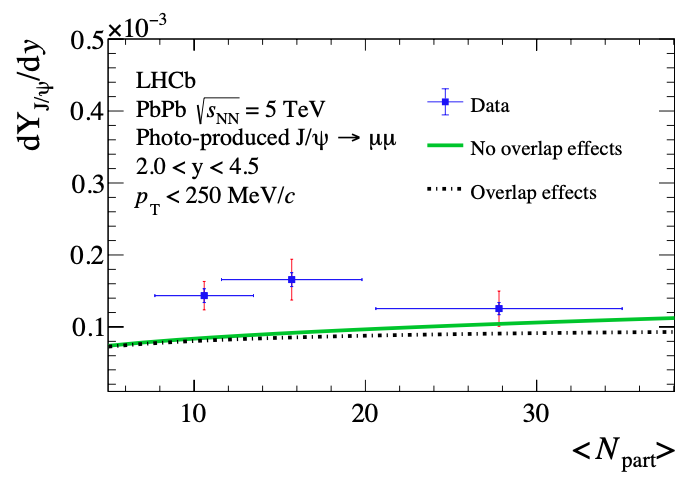}
\end{subfigure} \\
\begin{subfigure}{.485\linewidth}		   							   		\includegraphics[width=1.0\linewidth]{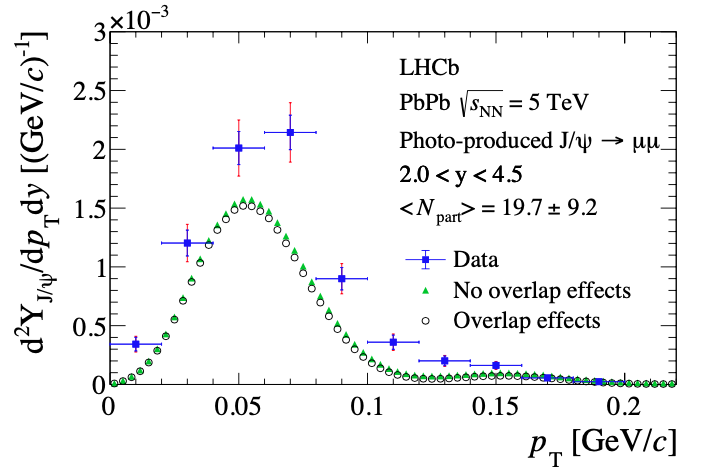}
\end{subfigure}
\caption{Measured signal yields for the $J/\varPsi$ calculated as a function of: (a) $y$, (b) $\left< N_{c} \right>$, and (c) double-differentially of $p_{T}$ and $y$. Also plotted are theoretical predictions, both with and without overlap effects. \cite{LHCb:2021hoq}}
\label{paper3}
\end{figure}

In peripheral collisions, the impact parameter is between one and two times the radius of the nucleus and, in contrast to UPCs, the hadron dissociates. This analysis \cite{LHCb:2021hoq} is useful as, previously, both the ALICE \cite{ALICE:2015mzu} and STAR \cite{STAR:2019yox} collaborations measured an excess with respect to expectations from purely hadronic $J/\varPsi$ production at very low $p_{T}$. It has been posited \cite{PhysRev.45.729} that this excess is due to photo-production caused by electromagnetic coherent interactions. These were previously thought to only occur in UPCs, so their presence in peripheral collisions provides an interesting opportunity to study the coherence of the interaction and examine the profile of the photon flux in peripheral $Pb-Pb$ collisions. Candidates are reconstructed through the dimuon decay channel. The $Pb-Pb$ sample is divided into intervals of $N_{c}$ (the number of clusters detected in the VELO) which is a proxy for the number of participating nucleons $N_{part}$ in the event. For each of the intervals, the mean number of participating nucleons is calculated. The photo-produced and hadronically produced $J/\varPsi$ mesons are disentangled through an unbinned fit to the dimuon $p_{T}$ spectrum. They are calculated as a function of rapidity $\left( y \right)$, as a function of the mean number of participating nucleons $\left< N_{part} \right>$, and double-differentially as a function of transverse momentum $\left( p_{T} \right)$. The results are shown in Figure \ref{paper3}. These results are the most precise to date, and support the hypothesis of coherent $J/\varPsi$ photo-production in peripheral hadronic collisions. Plotted also are theoretical predictions \cite{Zha:2017jch,Zha:2018jin}, both with and without overlap effects. The shape of the results is qualitatively described by the predictions, although a normalisation discrepancy is observed which could possibly hide an additional contribution.

\section{Measurement of prompt charged-particle production in $pp$ collisions at $\sqrt{s} = 13$ TeV}
\label{sec:paper4}
\begin{figure}[h]
\centering
\includegraphics[width=1.0\linewidth]{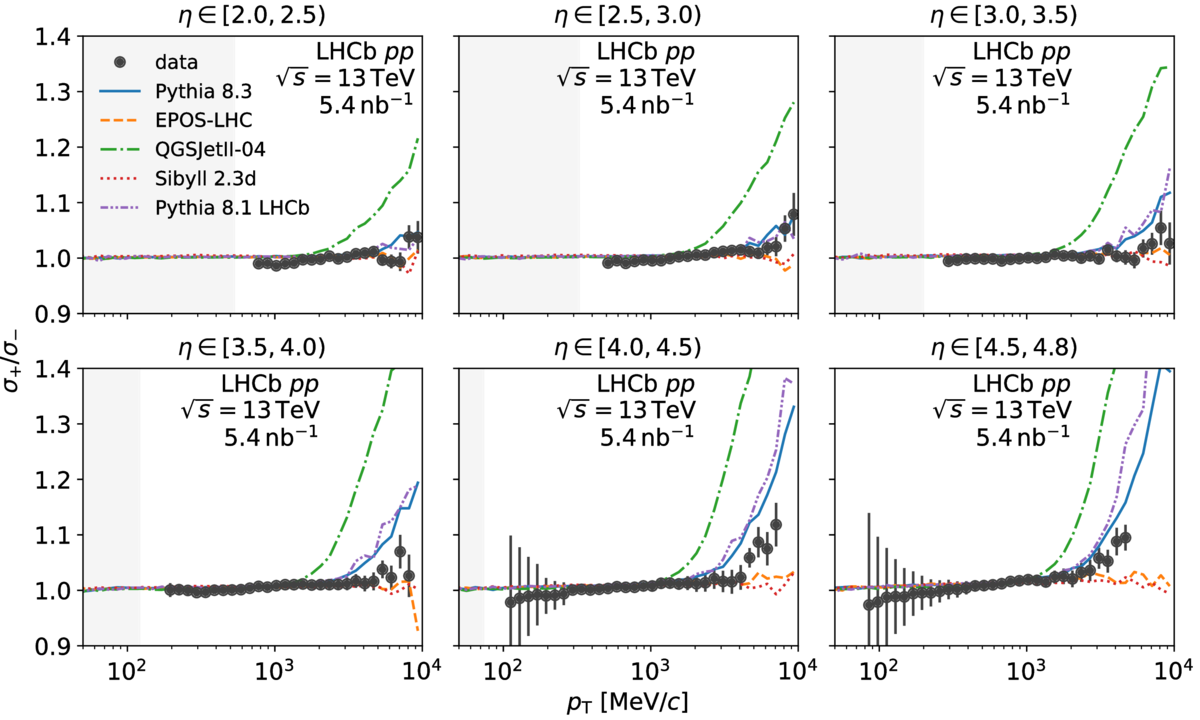}
\caption{Ratio of the differential cross sections for positively and negatively charged particles in $\eta$ intervals. Plotted for comparison are a number of model predictions. \cite{LHCb:2021abm}}
\label{paper4}
\end{figure}

Hadron production in inelastic high-energy proton-proton collisions is dominated by soft processes that cannot be described by perturbative-QCD. Instead, phenomenological models with experimental input are needed. Monte Carlo generators that implement these models are used at LHC experiments and in cosmic-ray research, where generators are used to simulate the interactions of ultra-relativistic nuclei within the atmosphere, inducing particle cascades known as air showers. There is an observed numerical excess of muons produced in actual air showers when compared to simulation, referred to as the muon puzzle \cite{Albrecht:2021cxw}. Precision measurements of light hadron production at the TeV scale in the forward region can be used to guide and constrain the models and extrapolate them to higher collision energies. A good proxy for light hadron production is the production of prompt\footnote{Prompt here means that the charged particles are long-lived and originate from the primary interaction.} long-lived charged particles. The precision achieved in this measurement \cite{LHCb:2021abm} is useful for an improved simulation of the underlying events in collisions at the LHC and of the interactions in the Earth's atmosphere that cause air showers. A small set of loose requirements is used to maintain high selection efficiency and the analysis uses an unbiased trigger. The plots shown in Figure \ref{paper4} show the ratio of the differential cross sections for positively and negatively charged particles in intervals of $\eta$, along with a number of model predictions. At high $\eta$ and high $p_{T}$, the production of positively charged particles increases as the initial state has a charge of +2, which transfers to the final state in the forward region. The models predict this to varying degrees, with the ratio best predicted by \textsc{Pythia 8.3} \cite{Sjostrand:2014zea}, although the onset of the increase is shifted towards lower $p_{T}$ values.

\section{Conclusion}
For each of the papers presented in these proceedings, the main results are summarised below.
\begin{itemize}
\item \textbf{CEP of $J/\varPsi$ and $\psi(2S)$ mesons in $pp$ collisions at $\sqrt{s} = 13$ TeV:} Photo-production cross sections found to be in good agreement with the JMRT NLO. For $J/\varPsi$ there is a deviation from a pure power-law extrapolation of H1 data. The $\psi(2S)$ results are consistent, although more data is required to make a critical comparison.
\item \textbf{Study of the coherent charmonium production in lead-lead ultra-peripheral collisions:} Measurements of differential cross section as a function of $y$ and $p_{T}$ for the $J/\varPsi$ and $\psi(2S)$ are found to be compatible with theoretical models.
\item \textbf{$J/\varPsi$ photo-production in Pb-Pb peripheral collisions at $\sqrt{s_{NN}} = 5$ TeV:} Most precise results yet. Supports the hypothesis of coherent $J/\varPsi$ photo-production in peripheral hadronic collisions. Results are qualitatively described by the theoretical prediction.
\item \textbf{Measurement of prompt charged-particle production in $pp$ collisions at $\sqrt{s} = 13$ TeV:} Results compared with predictions. Models mostly over estimate differential cross section. Ratio of the differential cross sections best predicted by \textsc{Pythia 8.3}.
\end{itemize} 

\section*{Acknowledgements}
\paragraph{Funding information}
The author would like to thank to organisers of the \emph{International Symposium  on Multiparticle Dynamics 2022} for subsiding his attendance.






\bibliography{ISMD2022_proceedings.bib}

\nolinenumbers

\end{document}